\pdfoutput=1
\documentclass[11pt,a4paper]{article}
\usepackage{a4wide,german,amsmath}
\usepackage[latin1]{inputenc}
\usepackage{graphicx}

\begin{document}

\title{Supersymmetrische Kandidaten f\"ur die Dunkle Materie}

\author{Frank Daniel Steffen\\
        Max-Planck-Institut für Physik, Föhringer Ring 6, 80805 München}

\date{}

\maketitle

\begin{abstract}
  Die teilchenphysikalische Identit\"at der Dunklen Materie ist eines
  der gro{\ss}en R\"atsel unseres Universums.
  Seine L\"osung kann mit einer bisher noch nicht nachgewiesenen
  fundamentalen Raumzeit-Symmetrie verkn\"upft sein: der
  Supersymmetrie.
  In vielen supersymmetrischen Erweiterungen des Standardmodells der
  Elementarteilchenphysik kann das leichteste supersymmetrische
  Teilchen nicht zerfallen und ist daher ein vielversprechender
  Kandidat f\"ur die Dunkle Materie.
  Das leichteste Neutralino, das bereits in dem minimalen
  supersymmetrischen Modell auftritt, kann als ein solcher Kandidat in
  der indirekten Suche, der direkten Suche und \"uber die Produktion
  an zuk\"unftigen Beschleunigern identifiziert werden.
  Auch das Gravitino, der Superpartner des Gravitons, liefert als
  m\"ogliches leichtestes Superteilchen eine m\"ogliche Erkl\"arung
  der Dunklen Materie.
  Es kann weder in der direkten oder der indirekten Suche nach der
  Dunklen Materie nachgewiesen noch direkt an Beschleunigern
  produziert werden.
  Die Untersuchung von Zerf\"allen langlebiger geladener Teilchen an
  zuk\"{u}nftigen Beschleunigern k\"onnte jedoch einen experimentellen
  Nachweis des Gravitinos erm\"oglichen.
  Die kommenden Experimente am CERN Large Hadron Collider k\"onnen so
  zu einem zentralen Schl\"ussel f\"ur das Verst\"andnis unseres
  Universums werden.
\\ \\

The identity of dark matter is one of the greatest puzzles of our
Universe.
Its solution may be associated with supersymmetry which is a
fundamental space-time symmetry that has not been verified
experimentally so far.
In many supersymmetric extensions of the Standard Model of particle
physics, the lightest supersymmetric particle cannot decay and is
hence a promising dark matter candidate.
The lightest neutralino, which appears already in the minimal
supersymmetric model, can be identified as such a candidate in
indirect and direct dark matter searches and at future colliders.
As the superpartner of the graviton, the gravitino is another
candidate for the lightest superparticle that provides a compelling
explanation of dark matter.
While it will neither be detected in indirect or direct searches nor be
produced directly at accelerators,
the analysis of late-decaying charged particles can allow for an
experimental identification of the gravitino at future accelerators.
In this way, the upcoming experiments at the CERN Large Hadron
Collider may become a key to the understanding of our Universe.
\end{abstract}

\section{Einleitung}

\begin{figure}[ht!]
\centering
\includegraphics[width=\textwidth]{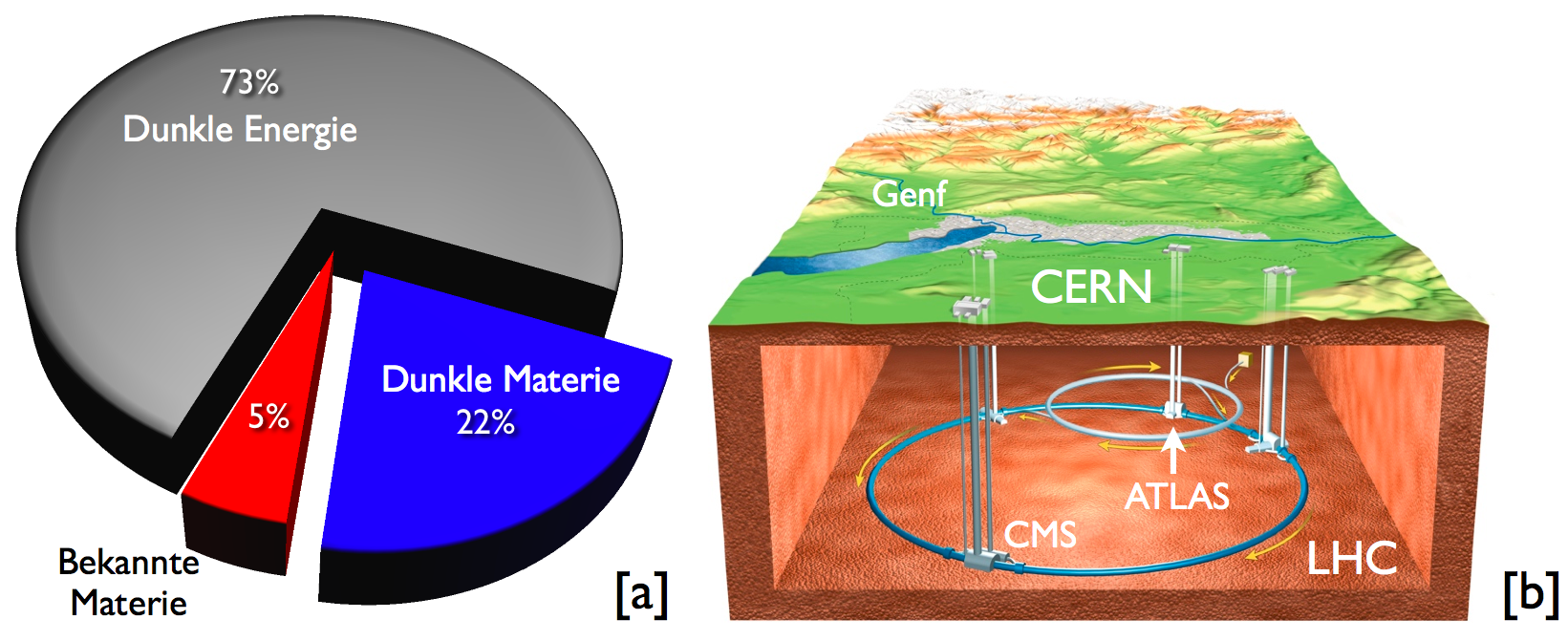}
\caption{(a) Der Gesamtenergieinhalt des Universums besteht nach heutigen Erkenntnissen zu ca.\ 73~\% aus Dunkler Energie, zu ca.\ 22~\% aus Dunkler Materie und nur zu ca.\ 5~\% aus den bekannten Teilchen, die bisher in teilchenphysikalischen Experimenten entdeckt und untersucht werden konnten. (Grafik:\ MPI f\"ur Physik) (b) Illustration des Teilchenbeschleunigers LHC am Forschungszentrum CERN in Genf. Der gro{\ss}e ringf\"ormige LHC Beschleuniger (blau) hat einen Umfang von 27~km und befindet sich etwa 100~m tief unter der Erdoberfl\"ache. An den Experimenten ATLAS (A Toroidal LHC ApparatuS) und CMS (Compact Muon Solenoid) k\"onnte bereits in den n\"achsten Jahren die Produktion und die Identifizierung des fundamentalen Bausteins der Dunklen Materie gelingen. (Grafik:\ CERN)}
\label{Fig:1}
\end{figure}

Zahlreiche astrophysikalische und kosmologische Beobachtungen deuten
darauf hin, dass unser Universum zu ca.\ 73\% aus Dunkler Energie und
zu ca.\ 22\% aus Dunkler Materie besteht. Diese Bestandteile des
Universums lassen sich nicht mit den Teilchen erkl\"aren, die bisher
in teilchenphysikalischen Experimenten entdeckt und untersucht werden
konnten.
Es liegen also nur ca.\ 5\% des Energieinhaltes unseres Universums in
Form der bekannten Teilchen vor (Abb.~\ref{Fig:1}a).
%
%
Mit zuk\"unftigen Teilchenbeschleunigern -- wie z.B.\ dem nahezu
fertiggestellten Large Hadron Collider (LHC) am Forschungszentrum CERN
in Genf (Abb.~\ref{Fig:1}b) -- k\"onnte es jedoch schon in den
n\"achsten Jahren gelingen, neue Teilchen und damit auch den
fundamentalen Baustein der Dunklen Materie zu produzieren und zu
identifizieren.
%
%

Die bisher entdeckten fundamentalen Teilchen und ihr Verhalten in
Experimenten werden sehr erfolgreich vom Standardmodell der
Elementarteilchenphysik beschrieben.
Basierend auf der Quantenfeldtheorie beschreibt dieses Modell drei der
vier fundamentalen Kr\"afte: die elektromagnetische Kraft, die
schwache Kraft und die starke Kraft.
Die vierte Kraft, die Gravitation, ist bei den experimentell
zug\"anglichen Energien sehr viel schw\"acher als die zuvor genannten.
Sie wird von Einsteins Allgemeiner Relativit\"atstheorie beschrieben,
deren Verkn\"upfung mit der Quantentheorie noch immer zu den
gr\"o{\ss}ten Herausforderungen der theoretischen Physik geh\"ort.

\begin{figure}[t!]
\centering
\includegraphics[width=\textwidth]{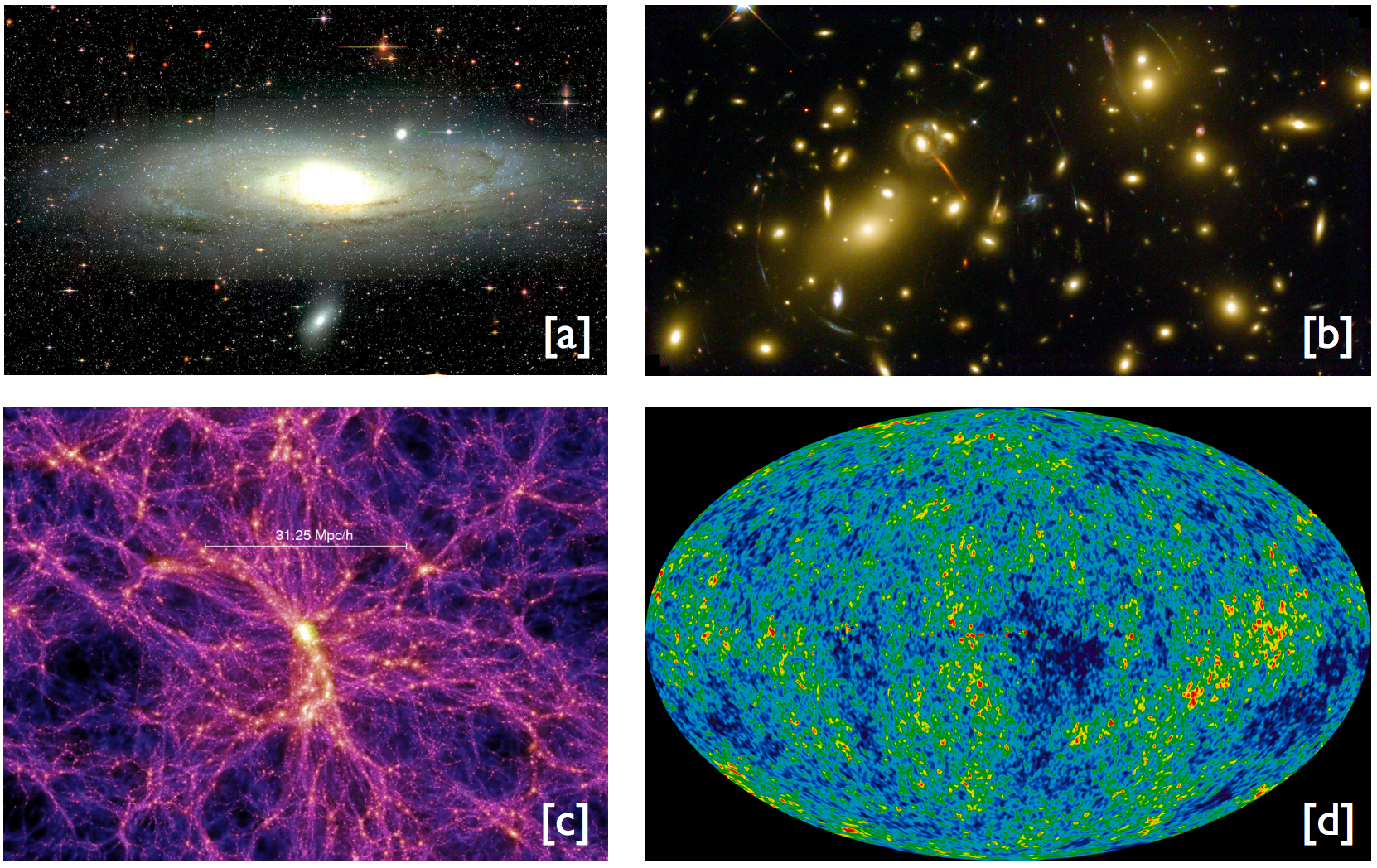}
\caption{Hinweise auf die Existenz der Dunklen Materie im Universum. (a) Die Spiralgalaxie M~31 (Andromeda Galaxie) hat ein Rotationsverhalten, das auf die Existenz der Dunklen Materie hindeutet. (Foto: Sloan Digital Sky Survey) (b) Der Galaxiehaufen Abell~2218 wirkt als Gravitationslinse und liefert sichelf\"ormige Bilder dahinterliegender Galaxien, die R\"uckschl\"usse auf die Menge der Dunklen Materie in dem Galaxiehaufen erlauben. (Foto: NASA, A.~Fruchter and the ERO Team) (c) Die Visualisierung der von der numerischen Millenium-Simulation berechneten heutigen Verteilung der Dunklen Materie im Universum zeigt eine Struktur, die sehr gut mit der beobachteten Verteilung der sichtbaren Galaxien \"ubereinstimmt. (Bild: MPI f\"ur Astrophysik) (d) Die von dem WMAP (Wilkinson Microwave Anisotropy Probe) Satelliten vermessenen winzigen Temperaturschwankungen in der kosmischen Mikrowellenstrahlung k\"onnen \"uberzeugend erkl\"art werden, 
%
%
wenn man annimmt, dass das Universum zu 73~\% aus Dunkler Energie, zu 22~\% aus Dunkler Materie und zu 5~\% aus den bekannten Teilchen besteht.
%
%
(Bild: NASA and the WMAP Science Team)}
\label{Fig:2}
\end{figure}

Beobachtungen der Gravitationsfelder von Galaxien und Galaxienhaufen
deuten auf die Existenz der Dunklen Materie hin.
Zum Beispiel liefern die hohe Rotationsgeschwindigkeit der sichtbaren
Materie in den \"au{\ss}eren Armen von Spiralgalaxien
(Abb.~\ref{Fig:2}a) oder die hohe Relativgeschwindigkeit von Galaxien
in Galaxienhaufen Hinweise auf Gravitationsfelder, die viel st\"arker
sind als die Gravitationsfelder, die man aufgrund der sichtbaren
gew\"ohnlichen Materie erwartet.
Die gro{\ss}e Materieansammlung in einem Galaxienhaufen kann auch als
Gravitationslinse wirken, die verzerrte Bilder dahinterliegender
Galaxien liefert (Abb.~\ref{Fig:2}b). Das Ausma{\ss} dieser
Verzerrungen l\"asst sich wiederum nur mit Gravitationsfeldern
erkl\"aren, deren St\"arke weit \"uber der von der sichtbaren Materie
erwarteten liegt.
Galaxien und Galaxienhaufen m\"ussen also zu einem Gro{\ss}teil aus
Materie bestehen, die Licht weder absorbiert noch emittiert: der
Dunklen Materie.

Die Dunkle Materie spielt nach heutigen Erkenntnissen insbesondere bei
der Bildung der gro{\ss}r\"aumigen Struktur im Universum
(Abb.~\ref{Fig:2}c) eine zentrale Rolle.
Bereits die winzigen Temperaturschwankungen in der kosmischen
Mikrowellenstrahlung (Abb.~\ref{Fig:2}d), die von Satelliten- und
Ballonexperimenten sehr genau vermessen werden, lassen sich
\"uberzeugend erkl\"aren, wenn man die Existenz der Dunklen Materie
annimmt.
Analysen dieser Temperaturschwankungen erm\"oglichen eine genaue
Bestimmung des Anteils der Dunklen Materie an der Gesamtenergie des
Universums.
Tats\"achlich charakterisieren die Temperaturschwankungen in der
kosmischen Mikrowellenstrahlung die Dichteschwankungen im fr\"uhen
Universum, die der Ausgangspunkt f\"ur die sp\"atere Bildung der
Galaxien und Galaxienhaufen waren.
Computersimulationen liefern -- von den Anfangsbedingungen aus den
Analysen der kosmischen Mikrowellenstrahlung ausgehend -- wertvolle
Einsichten in die kosmische Strukturbildung.
Unter der Annahme, dass die Dunkle Materie aus Teilchen mit
vernachl\"assigbaren Geschwindigkeiten besteht, zeigen diese
Simulationen ein Bild (Abb.~\ref{Fig:2}c), das sehr gut mit der
beobachteten Verteilung der Galaxien \"ubereinstimmt.

Da die Dunkle Materie sichtbares Licht und auch elektromagnetische
Strahlung mit anderen Wellenl\"angen weder absorbiert noch emittiert,
m\"ussen ihre Bestandteile elektrisch neutral sein.
Sie m\"ussen dar\"uber hinaus entweder stabil sein oder eine
Lebensdauer besitzen, die nicht weit unterhalb des Alters unseres
Universums liegen kann.
Mit einer k\"urzeren Lebensdauer w\"are ein Gro{\ss}teil der Dunklen
Materie heute bereits zerfallen. Dies widerlegen jedoch die auch noch
in der Milchstra{\ss}e und in nahegelegenen Spiralgalaxien
beobachteten Effekte der Dunklen Materie.

Unter den bekannten Teilchen des Standardmodells der
Elementarteilchenphysik besitzen nur die Neutrinos die
Grundeigenschaften der Dunklen Materie: Sie sind stabil, elektrisch
neutral und unterliegen allein der schwachen Kraft und der
Gravitation.
Die erst vor wenigen Jahren klar nachgewiesenen so genannten
Neutrinooszillationen zeigen dar\"uber hinaus, dass Neutrinos eine
Masse besitzen.
Diese Masse ist jedoch so klein, dass die Geschwindigkeit der
Neutrinos im jungen Universum und auch sp\"ater noch sehr hoch gewesen
sein muss.
Unter der Annahme, dass die Dunkle Materie aus den etablierten
Neutrinos besteht, ist es aufgrund dieser hohen Geschwindigkeiten
nicht m\"oglich, die Bildung und die Existenz der beobachteten
Strukturen in unserem Universum zu verstehen.
Im Standardmodell der Elementarteilchenphysik existiert somit kein
Teilchen, das die Dunkle Materie in \"Ubereinstimmung mit den
Beobachtungen erkl\"aren kann.

\section{Supersymmetrische Erweiterungen des Standardmodells der Elementarteilchenphysik}

Berechnet man die St\"arke der Standardmodell-Kr\"afte f\"ur Energien,
die viele Gr\"o{\ss}enordnungen oberhalb jener Energien liegen, die an
Teilchenbeschleunigern erreicht werden, dann findet man, dass sich die
unterschiedlichen Kopplungsst\"arken mit zunehmender Energie einem
gemeinsamen Wert ann\"ahern.
Dieses Verhalten kann ein Hinweis f\"ur die Vereinigung der drei
Standardmodell-Kr\"afte zu einer \"ubergeordneten Kraft sein.
Nimmt man jedoch die G\"ultigkeit des Standardmodells bis zu der
Vereinigungsenergieskala an, dann begegnet man dem so genannten
Hierarchieproblem:
Es stellt sich heraus, dass die aufgrund von Quanteneffekten erwartete
`nat\"urliche' Gr\"o{\ss}enordnung der Masse des so genannten
Higgs-Teilchens weit oberhalb des Bereiches liegt, den man indirekt in
Pr\"azisionsrechnungen aus den Daten von Experimenten an
Teilchenbeschleunigern erh\"alt. Solche Pr\"azisionsrechnungen
geh\"oren zu den Forschungsaktivit\"aten am MPI f\"ur Physik.

Im Standardmodell ist das Higgs-Teilchen f\"ur die Massen der
fundamentalen Teilchen verantwortlich und somit ein zentraler
Baustein.
Bisher konnte es noch nicht direkt beobachtet werden.
Es ist daher davon auszugehen, dass die f\"ur seine Produktion
notwendige Energie an den bisherigen Teilchenbeschleunigern nicht
erreicht werden konnte.
Der noch ausstehende direkte Nachweis des Higgs-Teilchens ist daher
mit ein Hauptgrund f\"ur den Bau des LHC am Forschungszentrum CERN in
Genf.
Dieser Teilchenbeschleuniger wird in den kommenden Jahren in einen
bisher in Labor-Experimenten unerreichten Energiebereich
vorsto{\ss}en.

Eine besonders elegante L\"osung des Hierarchieproblems ergibt sich in
supersymmetrischen Erweiterungen des Standardmodells.
Die Supersymmetrie ist eine fundamentale Raumzeit-Symmetrie zwischen
Elementarteilchen mit unterschiedlichem Spin, d.h.\ zwischen den
Materieteilchen und den die Kr\"afte vermittelnden Austauschteilchen,
die die Bausteine der zugrunde liegenden Quantenfeldtheorie
darstellen.
Sollte diese Symmetrie in der Natur realisiert sein, dann muss es mehr
als ein Higgs-Teilchen geben, und jedes der etablierten
Standardmodell-Teilchen muss einen supersymmetrischen Partner
besitzen.
Die Quanteneffekte dieser neuen Teilchen kompensieren im Falle der
Higgs-Masse die Quanteneffekte der Standardmodell-Teilchen. So kann
dann diese Masse auf nat\"urliche Weise in dem von den
Pr\"azisionsbetrachtungen erwarteten Bereich liegen.

Interessanterweise l\"ost die Supersymmetrie nicht nur das
Hierarchieproblem.
Aufgrund der von der Supersymmetrie postulierten neuen Teilchen
verhalten sich die Kopplungsst\"arken der drei Standardmodell-Kr\"afte
bei der Extrapolation zu hohen Energien so, dass sie sich am
Vereinigungspunkt tats\"achlich in einem Punkt treffen, was im Rahmen
des nicht-super\-symmetrischen Standardmodells nicht der Fall ist.
Dieses Verhalten untermauert die Hypothese der Vereinigung der drei
Standardmodell-Kr\"afte zu einer \"ubergeordneten Kraft.
Ebenso wie die Higgs-Teilchen wurden die Superpartner der
Standardmodell-Teilchen bisher nur aus theoretischen \"Uberlegungen
vorhergesagt.
Experimentell konnte die Existenz dieser Teilchen noch nicht
nachgewiesen werden.
Doch auch hier wird auf Entdeckungen am LHC gehofft.

\section{Supersymmetrische Dunkle Materie}

Sollte die Supersymmetrie in der Natur realisiert sein, dann
wird -- auch aufgrund der beobachteten Stabilit\"at des Protons -- davon
ausgegangen, dass supersymmetrische Prozesse eine diskrete Symmetrie,
die so genannte R-Parit\"at, respektieren.
Die dazugeh\"orige Quantenzahl unterscheidet zwischen den
Standardmodell- und den Higgs-Teilchen, die eine {\em gerade}
R-Parit\"at ($+1$) besitzen, und ihren Superpartnern, die eine {\em
  ungerade} R-Parit\"at ($-1$) besitzen.
Ein Prozess erh\"alt nur dann die R-Parit\"at, wenn das Produkt der
R-Parit\"aten der Teilchen im Anfangszustand gleich dem der Teilchen
im Endzustand ist.

Aus der geforderten Erhaltung der R-Parit\"at folgt, dass aus dem
Zerfall eines Superpartners immer ein weiterer Superpartner
hervorgehen muss.
Da die Massen von Zerfallsprodukten aufgrund von
Energie-Impuls-Erhaltung immer unterhalb der Masse des zerfallenden
Teilchens liegen m\"ussen, impliziert dies, dass der leichteste
Superpartner -- auch wenn er deutlich schwerer als die etablierten
Standardmodell-Teilchen ist -- bei erhaltener R-Parit\"at nicht
zerfallen kann.
Der leichteste Superpartner wird also stabil sein.
Sollte er im fr\"uhen Universum produziert worden sein, 
dann k\"onnte er noch heute in gro{\ss}en Mengen vorhanden sein.
Ein elektrisch neutraler leichtester Superpartner ist somit
ein vielversprechender Kandidat f\"ur die Dunkle Materie.

Tats\"achlich liefern supersymmetrische Erweiterungen des
Standardmodells auf sehr nat\"urliche Weise solche
Dunkle-Materie-Kandidaten.
Hierzu geh\"oren u.a.\ das leichteste Neutralino und das Gravitino.
Im Folgenden werden diese noch hypothetischen Teilchen, ihre
Eigenschaften und Nachweism\"oglichkeiten n\"aher beschrieben.

\subsection{Das leichteste Neutralino}

Die Neutralinos sind Superteilchen, die aus Mischungen der elektrisch
neutralen Superpartner der Higgs-Teilchen, des Photons und des so
genannten Z-Bosons bestehen.
Wie die Neutrinos unterliegen sie nur der schwachen Kraft und der
Gravitation.
Sollten die Neutralinos existieren, dann m\"ussen sie -- da sie noch
nicht an Teilchenbeschleunigern nachgewiesen werden konnten -- um
Gr\"o{\ss}enordnungen schwerer sein als die Neutrinos.
Dies motiviert die Klassifizierung eines Neutralinos als ein Weakly
Interacting Massive Particle (WIMP).

Die Temperaturen direkt nach dem Urknall k\"onnen um weitere
Gr\"o{\ss}enordnungen h\"oher gewesen sein als die Masse der
Neutralinos. Bei solch hohen Temperaturen konnten alle
Standardmodell-Teilchen, die Higgs-Teilchen und ihre Superpartner
effizient produziert werden.
Die primordialen Dichten dieser Teilchen k\"onnen daher so hoch
gewesen sein, dass ein thermisches Gleichgewicht von Produktions- und
Vernichtungsprozessen vorherrschte.  Die H\"aufigkeit jeder einzelnen
Teilchensorte war dann in dieser hei{\ss}en Epoche vergleichbar mit
der der Photonen.
Mit der Ausdehnung des Universums nimmt jedoch die Temperatur ab.  Bei
Temperaturen unterhalb der Masse eines Teilchen wird die Dichte dieser
Teilchensorte sehr schnell sehr klein. Es ist nicht mehr genug
thermische Energie vorhanden, um diese Teilchen weiter produzieren zu
k\"onnen. Je schwerer ein Teilchen ist, das sich im thermischen
Gleichgewicht befindet, umso fr\"uher nimmt seine Dichte rapide ab.
Auch die Dichte des leichtesten Neutralinos nimmt stark ab bis zu dem
Zeitpunkt, an dem die Temperatur ein Bruchteil seiner Masse betr\"agt.
An diesem Punkt ist die Dichte der Neutralinos und der anderen
Superpartner so gering, dass ein Neutralino praktisch keinen
Reaktionspartner mehr f\"ur einen Vernichtungsprozess findet.
Solch ein weiterer Superpartner ist notwendig, da aufgrund der
Erhaltung der R-Parit\"at Superpartner nur paarweise produziert oder
vernichtet werden k\"onnen.
Ist die Wahrscheinlichkeit f\"ur Neutralino-Vernichtungsprozesse
vernachl\"assigbar, dann entkoppelt das leichteste Neutralino vom
thermischen Plasma, so dass die Anzahl dieser Teilchensorte im
Universum sich (nahezu) nicht mehr \"andert.

Mit Computerprogrammen kann man heute sehr genau die Entkopplung des
leichtesten Neutralinos im fr\"uhen Universum und daraus seine heutige
H\"aufigkeit berechnen.
Die Ergebnisse zeigen, dass die Dichte der Neutralinos heute
tats\"achlich mit der beobachteten Dichte der Dunklen Materie
\"ubereinstimmen kann.
Dar\"uber hinaus ist die Geschwindigkeit der aus dem thermischen
Gleichgewicht entkoppelten Neutralinos vernachl\"assigbar.
Unter der Annahme, dass das leichteste Neutralino der fundamentale
Baustein der Dunklen Materie ist, l\"asst sich somit die Bildung und
die Existenz der beobachteten Strukturen in unserem Universum gut
verstehen.
Es k\"onnte also eine Anh\"aufung der leichtesten Neutralinos sein,
die den Gro{\ss}teil der Masse in Galaxien und Galaxienhaufen
ausmacht.

Drei zueinander komplement\"are Methoden werden verfolgt, um dieses
theoretisch ansprechende Bild experimentell zu verifizieren:
die indirekte Suche, die direkte Suche und die Produktion an Beschleunigern.
\begin{figure}[ht!]
\centering
\includegraphics[width=\textwidth]{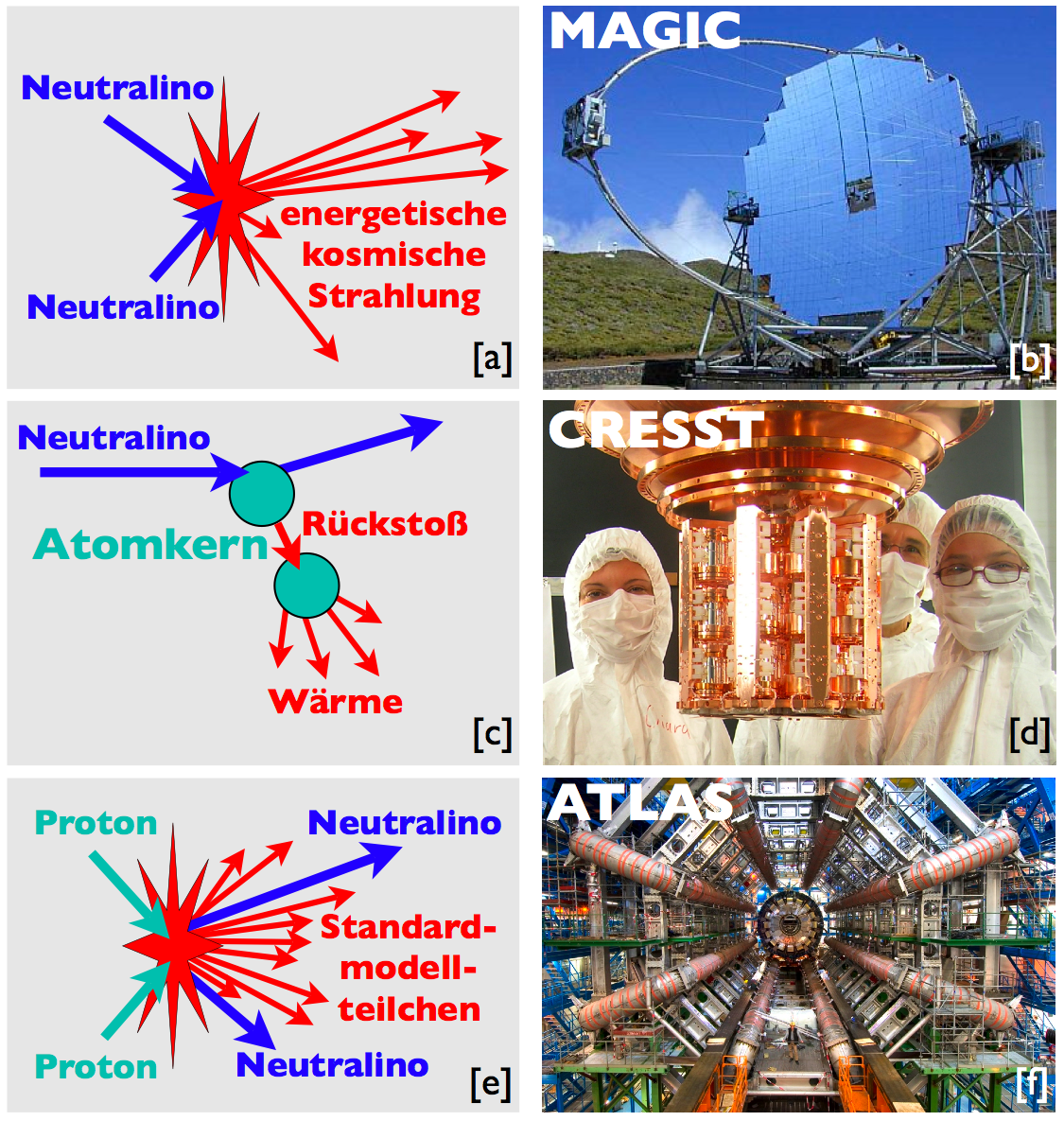}
\caption{Nachweism\"oglichkeiten von Neutralinos. (a) Treffen zwei Neutralinos aufeinander, dann werden diese in energetische kosmische Strahlung umgewandelt. (Grafik: MPI f\"ur Physik) (b) Diese sollte z.B.\ mit dem Teleskop MAGIC beobachtbar sein. (Foto: MPI f\"ur Physik) (c) St\"o{\ss}t ein Neutralino auf einen Atomkern, dann erf\"ahrt dieser einen R\"ucksto{\ss}, der das umliegende Material erw\"armt. (Grafik: MPI f\"ur Physik) (d) Das CRESST Experiment sucht nach diesen Ereignissen in tiefgek\"uhlten, in Modulen angeordneten Kristallen. (Foto: MPI f\"ur Physik) (e) Neutralinos sollten paarweise z.B. in den Proton-Proton Kollisionen am LHC produziert werden k\"onnen. (Grafik: MPI f\"ur Physik) (f) Zur Identifikation der unsichtbaren Neutralinos m\"ussen Teilchendetektoren, wie z.B.\ der nahezu fertiggestellte ATLAS Detektor, die zus\"atzlich produzierten Standardmodell-Teilchen extrem genau vermessen. (Foto: CERN)}
\label{Fig:3}
\end{figure}

Bei der indirekten Suche wird nach Signalen aus
Neutralino-Vernichtungsprozessen gesucht. Man erwartet, dass in
Regionen des Universums, in denen die Konzentration der Dunklen
Materie \"uberdurchschnittlich hoch ist, wie z.B.\ in Galaxien, noch
heute gelegentlich Neutralino-Vernichtungsprozesse stattfinden. Obwohl
diese Prozesse keinen gro{\ss}en Einfluss auf die
Neutralino-H\"aufigkeit haben, sollten in dieser Reaktion
Standardmodell-Teilchen emittiert werden.
Diese Standardmodell-Teilchen k\"onnen eine hohe Energie besitzen und
somit zu energetischen kosmischen Strahlen f\"uhren
(Abb.~\ref{Fig:3}a), die in Erdn\"ahe und auf der Erde beobachtbar
sein sollten.
M\"oglicherweise sind solche kosmischen Strahlen bereits von dem
satellitenbasierten Experiment EGRET (Energetic Gamma Ray Experiment
Telescope) beobachtet worden.
Hier besteht jedoch die Herausforderung, bei der Interpretation des
beobachteten Spektrums, Signale von Neutralino-Vernichtungsprozessen,
die auch von der genauen Verteilung der unsichtbaren Dunklen Materie
abh\"angen, aus dem komplizierten Hintergrund von anderen Quellen
hochenergetischer Strahlen herauszufiltern.
Das Weltraumteleskop GLAST (Gamma-Ray Large Area Space Telescope) wird
hierzu in Zukunft neue Daten liefern.
Auch in so genannten Cherenkov-Teleskopen, wie z.B.\ H.E.S.S.\ (High
Energy Stereoscopic System) oder $\mbox{MAGIC}$ (Major Atmosspheric
Gamma Imaging Cherenkov), an dem das MPI f\"ur Physik beteiligt ist
(Abb.~\ref{Fig:3}b), sollten Signale aus
Neutralino-Vernichtungsprozessen beobachtbar sein.

Bei der direkten Suche wird nach Signalen von Neutralinos aus deren
Kollision mit Atomkernen gesucht. Als Dunkle Materie sollten
Neutralinos auch die Erde in solch einer H\"aufigkeit umgeben, dass
sie trotz ihrer schwachen Wechselwirkungen gelegentlich auf Atomkerne
stossen. Da die Masse der Neutralinos so gro{\ss} ist, erf\"ahrt der
Kern hierbei einen deutlichen R\"ucksto{\ss}, der zu einer minimalen
Temperaturerh\"ohung in dem den Kern umgebenden Material f\"uhrt
(Abb.~\ref{Fig:3}c).
Die Herausforderung ist, die von den Neutralinos verursachten
R\"uckst\"o{\ss}e von denen zu unterscheiden, die durch andere
Teilchen verursacht werden.
Vielversprechende Methoden hierzu finden z.B.\ in den Experimenten
CDMS (Cryogenic Dark Matter Search), CRESST (Cryogenic Rare Event
Search with Superconducting Thermometers) und EDELWEISS (Exp\'erience
pour DEtecter Les WIMPs En Site Souterrain) Verwendung.
Zur Abschirmung von st\"orenden Standardmodell-Teilchen befinden sich
diese Experimente unter mehreren hundert Metern von Gestein, die
allerdings von den schwach wechselwirkenden Neutralinos problemlos
durchdrungen werden k\"onnen.
Das MPI f\"ur Physik ist an dem Experiment CRESST, das sich im
Gran-Sasso-Untergrundlabor befindet, beteiligt (Abb.~\ref{Fig:3}d).

Schwere, nur schwach wechselwirkende Teilchen werden seit ca.\ 20
Jahren mit Hilfe von Teilchenbeschleunigern produziert und untersucht.
Kurzlebige Z-Bosonen, die die bisher schwersten beobachteten Teilchen
sind, die allein der schwachen Kraft und der Gravitation unterliegen,
konnten so bereits zahlreich produziert werden.
Die Wahrscheinlichkeit, Neutralinos zu produzieren, sollte \"ahnlich
sein, sofern die Energien an den Teilchenbeschleunigern hoch genug
sind.
Somit ist es gut m\"oglich, dass bereits in den n\"achsten Jahren
Neutralinos in den Proton--Proton Kollisionen am LHC erzeugt werden
k\"onnen (Abb.~\ref{Fig:3}e).
W\"ahrend ein Z-Boson \"uber seinen Zerfall und die daraus
resultierenden sichtbaren Zerfallsprodukte nachgewiesen werden kann,
hinterl\"asst ein stabiles leichtestes Neutralino keine Spur in den
Teilchendetektoren und kann daher nur mit Hilfe der
Energie-Impuls-Erhaltung nachgewiesen werden.
Dies erfordert eine extrem genaue Vermessung der Spuren der
zus\"atzlich produzierten Standardmodell-Teilchen, die auch an dem
noch im Aufbau befindlichen ATLAS Experiment (Abb.~\ref{Fig:3}f) unter
Beteiligung des MPI f\"ur Physik erreicht werden soll.
Von theoretischer Seite muss dar\"uber hinaus der Mechanismus, der zur
Neutralino-Produktion am Beschleuniger f\"uhrt, sehr gut verstanden
sein.
Dies ist ein Gegenstand aktueller Forschung auch am MPI f\"ur Physik.

\subsection{Das Gravitino}

Das Gravitino ist der Superpartner des Gravitons, das das
Austauschteilchen der Gravitation und als solches nicht Teil des
Standardmodells der Elementarteilchenphysik ist.
Die Wechselwirkungen des Gravitinos sind, wie die des Gravitons,
sensitiv auf die Energien der Interaktionspartner.
Die St\"arke dieser Wechselwirkungen ist dar\"uber hinaus durch die
sehr kleine Newtonsche Gravitationskonstante gegeben.
Tats\"achlich ist die Wahrscheinlichkeit f\"ur die Interaktion eines
Gravitinos im Labor so klein, dass ein Gravitino an
Teilchenbeschleunigern nicht direkt produziert werden kann, selbst
wenn seine Masse in dem Energiebereich liegt, der bereits zug\"anglich
ist.
Die Tatsache, dass noch kein Gravitino an Beschleunigern nachgewiesen
werden konnte, l\"asst daher keine R\"uckschl\"usse auf die
Gravitino-Masse zu.
Wird das Gravitino als Dunkle-Materie-Kandidat betrachtet, dann wird
jedoch angenommen, dass es der leichteste existierende Superpartner
und somit leichter als die Standardmodell-Superpartner, zu denen auch
das oben diskutierte leichteste Neutralino geh\"ort, ist.
Aufgrund der extrem unterdr\"uckten Wechselwirkungen und der
unbekannten Masse kann man das Gravitino als ein Extremely Weakly
Interacting Particle klassifizieren.

Hier muss noch betont werden, dass das Gravitino -- im Gegensatz zum
masselosen Graviton -- nur deshalb eine Masse besitzen kann, da bei
niedrigen Energien die Supersymmetrie nicht als eine exakte, sondern
nur als eine so genannte spontan gebrochene Symmetrie vorliegen kann.
Im Falle einer exakten Supersymmetrie m\"ussten die Massen der
Superpartner identisch mit denen der zugeh\"origen
Standardmodell-Teilchen sein. Dies kann jedoch ausgeschlossen werden,
da Superpartner an Teilchenbeschleunigern bisher noch nicht beobachtet
werden konnten. Tats\"achlich ist der Wert der Gravitino-Masse direkt
mit der so genannten Brechungsskala der Supersymmetrie, also der
Energieskala oberhalb derer die Supersymmetrie wieder als eine
perfekte Symmetrie betrachtet werden kann, verkn\"upft und daher von
fundamentaler Bedeutung in supersymmetrischen Theorien.

Die St\"arke der Gravitino-Wechselwirkungen w\"achst an, wenn die
Energien der Interaktionspartner zunehmen.  In den ersten Momenten
unseres Universums k\"onnen die Temperaturen und damit die Energien
der Standardmodell-Teilchen, der Higgs-Teilchen und ihrer Superpartner
so gro{\ss} gewesen sein, dass eine effiziente Produktion von
Gravitinos m\"oglich war.
Gravitino-Vernichtungsprozesse sind typischerweise vernachl\"assigbar,
so dass ein thermisches Gleichgewicht f\"ur Gravitinos nicht vorliegt.
Die Dichte der thermisch produzierten Gravitinos muss daher mit
Methoden der Quantenfeldtheorie bei endlichen Temperaturen berechnet
werden.
Solche Rechnungen sind in den letzten Monaten auch am MPI f\"ur Physik
durchgef\"uhrt worden.

Zur gesamten Gravitino-Dichte tr\"agt auch die nicht-thermische
Produktion bei, die deutlich sp\"ater -- also bei viel kleineren
Temperaturen -- als die thermische stattfand.
In einem Szenario, in dem das Gravitino der leichteste Superpartner
ist, sind die anderen Superpartner nicht stabil.
Diese anderen Superpartner k\"onnen auch bei niedrigen Temperaturen
stets in ein Gravitino zerfallen.
Da allerdings bei niedrigen Temperaturen sogar die schwache Kraft um
viele Gr\"o{\ss}enordnungen st\"arker als die die
Gravitino-Wechselwirkung bestimmende Gravitation ist, zerfallen die
schwereren Superpartner zuerst in den leichtesten
Standardmodell-Superpartner.
Dieser kann das leichteste Neutralino oder auch ein elektrisch
geladenes Teichen sein, wie z.B.\ der Superpartner des Tau-Leptons:
das so genannte Stau.
Die Lebensdauer des leichtesten Standardmodell-Superpartners h\"angt
von seiner Masse und der Gravitino-Masse ab und kann in einem Bereich
von einigen Sekunden bis hin zu Jahren liegen.
Mit solch einer langen Lebensdauer verh\"alt sich der leichteste
Standardmodell-Superpartner im fr\"uhen Universum so, als w\"are er
stabil.
Seine Dichte vor dem Zerfall l\"asst sich daher, wie bereits oben
f\"ur das leichteste Neutralino beschrieben, berechnen.
Schlussendlich zerf\"allt jedoch jedes einzelne Teilchen dieser Sorte
in ein Gravitino.
Diese nicht-thermisch produzierten Gravitinos sind bei der Berechnung
der gesamten Gravitino-Dichte stets einzubeziehen.

W\"ahrend die thermisch produzierte Gravitino-Dichte von der
Gravitino-Masse und der Anfangstemperatur des fr\"uhen
strahlungsdominierten Universums abh\"angt, ist die nicht-thermisch
produzierte sensitiv auf das Massen-Spektrum der Superpartner, aber
typischerweise unabh\"angig von der obigen Anfangstemperatur.
Unter realistischen Annahmen f\"ur diese Anfangstemperatur, die
Gravitino-Masse und das Massen-Spektrum der
Standardmodell-Superpartner findet man, dass eine \"Ubereinstimmung
der Gravitino-Dichte mit der beobachteten Dunkle-Materie-Dichte
m\"oglich ist.

Sollte das Gravitino tats\"achlich der fundamentale Baustein der
Dunklen Materie sein, dann muss die Geschwindigkeit eines
Gro{\ss}teils der Gravitinos hinreichend klein sein, da ansonsten die
Bildung und die Existenz der beobachteten Strukturen in unserem
Universum nicht erkl\"art werden kann.
Diese Bedingung liefert Untergrenzen einerseits f\"ur die
Gravitino-Masse, die die Geschwindigkeit der thermisch produzierten
Gravitinos bestimmt, und anderer\-seits f\"ur die Masse des
leichtesten Standardmodell-Superpartners, die die Geschwindigkeit der
nicht-thermisch produzierten Gravitinos bestimmt.
Einschr\"ankungen an die Gravitino-Masse sind von besonderem
Interesse, da sie, wie oben beschrieben, direkt mit der Brechungsskala
der Supersymmetrie verkn\"upft ist.

\begin{figure}[t!]
\centering
\includegraphics[width=\textwidth]{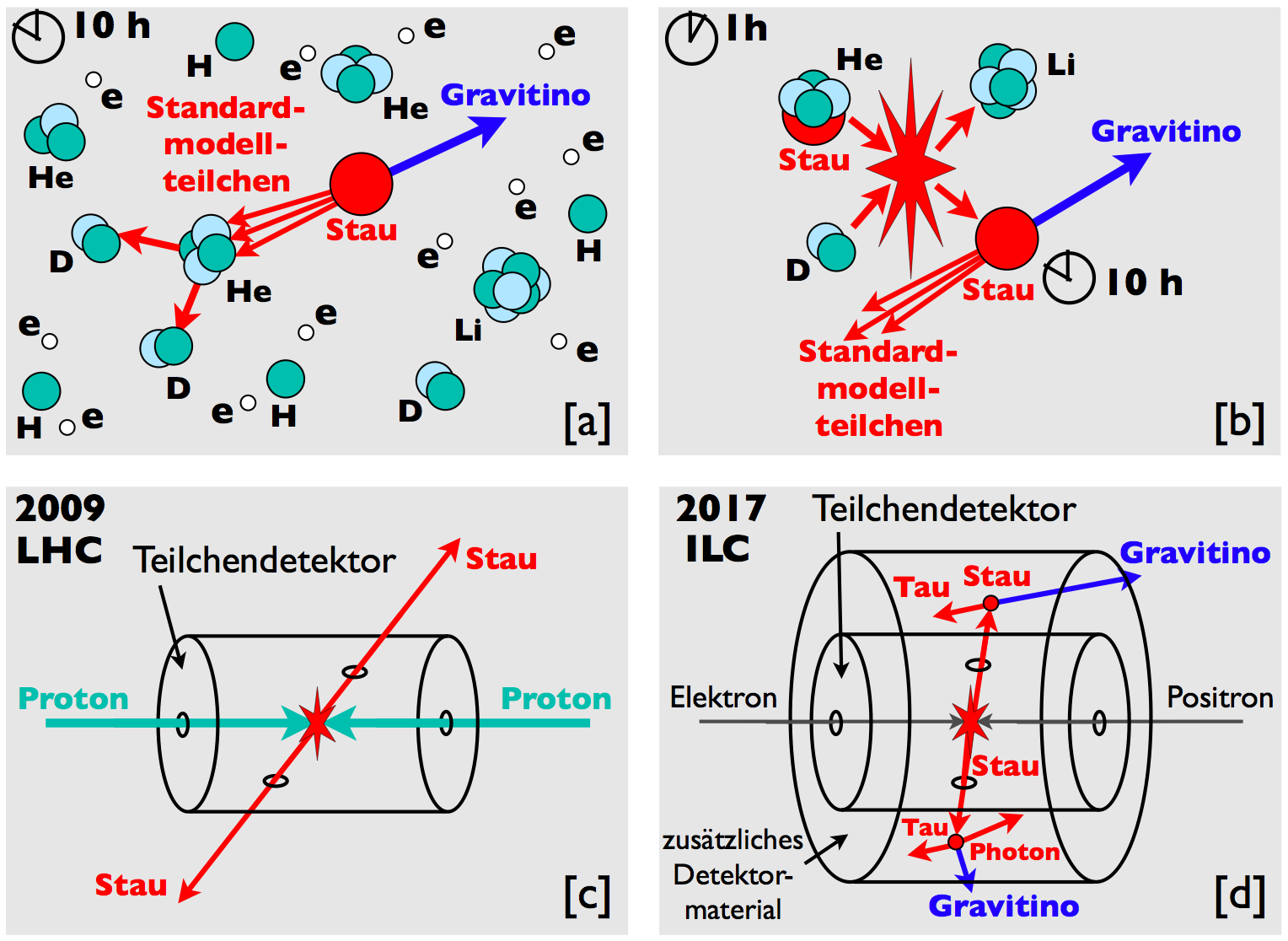}
\caption{
  (a)~Das Stau kann als leichtester Standardmodell-Superpartner eine
  Lebensdauer von zehn Stunden besitzen und im fr\"uhen Universum nach
  der Synthese der Atomkerne Deuterium, Helium und Lithium zerfallen.
  Die hierbei emittierten Standardmodell-Teilchen k\"onnen einen
  Helium-Kern in zwei Deuterium-Kerne zerlegen.
  (b) Vor seinem Zerfall kann ein Stau einen gebundenen Zustand mit
  einem Helium-Kern eingehen und so die primordiale Produktion von
  Lithium erleichtern.
  (c) Staus k\"onnten in den Proton-Proton Kollisionen am LHC
  produziert werden und bereits mit einer Lebensdauer von einer
  Sekunde wie scheinbar stabile Teilchen den Teilchendetektor gut
  sichtbar durchfliegen.
  (d) Es ist denkbar, dass in den Elektron-Positron Kollisionen an dem
  geplanten n\"achsten Linearbeschleuniger, dem International Linear
  Collider (ILC), langsame Staus, die noch im Teilchendetektor oder in
  zus\"atzlichem Detektormaterial stecken bleiben, produziert werden
  k\"onnen.  Die Untersuchung der darauffolgenden Zerf\"alle k\"onnte
  dann die Identifikation des Gravitinos als leichtester Superpartner
  und m\"ogliches Dunkle-Materie-Teilchen erlauben.  (Grafiken: MPI
  f\"ur Physik)}
\label{Fig:4}
\end{figure}

In Szenarien, in denen das Gravitino der leichteste Superpartner ist,
liefern die beobachteten H\"aufigkeiten der leichten Atomkerne Helium,
Deuterium und Lithium in unserem Universum weitere kosmologische
Einschr\"ankungen.
Diese leichten Elemente wurden bereits in der so genannten
primordialen Nukleosynthese produziert, die einsetzte, als das
Universum etwa eine Sekunde alt war.
Man glaubt heute, die primordiale Nukleosynthese weitgehend verstanden
zu haben.  Sie kann von Computerprogrammen simuliert werden. Hierbei
findet man, dass die berechneten primordialen Elementh\"aufigkeiten
gut mit den Beobachtungen \"ubereinstimmen.
Wenn nun der leichteste Standardmodell-Superpartner w\"ahrend oder
nach der primordialen Nukleosynthese zerf\"allt, dann werden neben dem
Gravitino auch Standardmodell-Teilchen emittiert, die mit den
produzierten leichten Atomkernen wechselwirken und dabei deren
H\"aufigkeit beeinflussen k\"onnen (Abb.~\ref{Fig:4}a).
Die Computersimulationen der primordialen Nukleosynthese liefern daher
zusammen mit den beobachteten Elementh\"aufigkeiten Obergrenzen f\"ur
die Emission von Standardmodell-Teilchen in Zerf\"allen des
leichtesten Standardmodell-Superpartners.
Sollte der leichteste Standardmodell-Superpartner eine elektrische
Ladung tragen, wie z.B.\ das Stau, dann kann dieser vor dem Zerfall
einen gebundenen Zustand mit den bereits produzierten positiv
geladenen Atomkernen eingehen und auch so Prozesse der primordialen
Nukleosynthese beeinflussen (Abb.~\ref{Fig:4}b).  Hieraus folgen
Obergrenzen f\"ur die H\"aufigkeit eines negativ geladenen leichtesten
Standardmodell-Superpartners.
Mit den diskutierten Obergrenzen l\"asst sich entscheiden, 
ob eine bestimmte supersymmetrische Erweiterung des Standardmodells, 
in der das Gravitino die Dunkle Materie liefern k\"onnte,
kosmologisch erlaubt ist.

Die experimentellen Nachweism\"oglichkeiten eines Gravitinos als
Dunkle-Materie-Kandidat sind st\"arker eingeschr\"ankt als die im
Falle des leichtesten Neutralinos.  Aufgrund der extrem
unterdr\"uckten Wechselwirkungen scheint es in absehbarer Zeit mit den
derzeit denkbaren Technologien nicht m\"oglich zu sein, Gravitinos in
der indirekten oder der direkten Suche, die oben f\"ur Neutralinos
beschrieben sind, experimentell nachzuweisen.
Auch die direkte Produktion von Gravitinos an Beschleunigern ist
aufgrund der geringen St\"arke der Wechselwirkungen sehr stark
unterdr\"uckt.
Dennoch gibt es Szenarien, in denen der experimentelle Nachweis
von Gravitinos an zuk\"unftigen Beschleunigern gelingen kann.

Mit dem Gravitino als leichtester Superpartner kann der leichteste
Standardmodell-Super\-partner, wie bereits oben erw\"ahnt, elektrisch
geladen sein. Dieses geladene Teilchen kann dar\"uber hinaus eine
Lebensdauer besitzen, die weit oberhalb von einer Sekunde liegt.
Sollte dieser leichteste Standardmodell-Superpartner an einem
zuk\"unftigen Beschleuniger produziert werden, dann w\"urde er als ein
langlebiges geladenes Teilchen eine gut sichtbare Spur in den
Teilchendetektoren hinterlassen.
Die Lebensdauer kann so gro{\ss} sein, dass dieses Teilchen nicht von
einem stabilen unterschieden werden kann, da seine Zerf\"alle erst
au{\ss}erhalb des Detektorvolumens stattfinden und so der Beobachtung
entgehen (Abb.~\ref{Fig:4}c).
Aufgrund der strengen kosmologischen Grenzen f\"ur die Existenz eines
schweren stabilen geladenen Teilchens k\"onnen Signaturen solcher
scheinbar stabiler leichtester Standardmodell-Superpartner bereits auf
die Existenz des Gravitinos oder eines anderen extrem schwach
wechselwirkenden leichtesten Superpartners hindeuten.

Da ein schweres geladenes langlebiges Teilchen auf seinem Weg durch
den Teilchendetektor Energie verliert, ist sogar denkbar, dass solch
ein leichtester Standardmodell-Superpartner, wenn er mit einer
geringen Anfangsgeschwindigkeit produziert wird, im Detektorvolumen
stecken bleiben und dort bis zu seinem Zerfall verharren kann. Die
Analyse dieses Zerfalls kann dann m\"oglich sein und zu einem
experimentellen Nachweis der Gravitinos f\"uhren (Abb.~\ref{Fig:4}d).

Die Theorie, die als Supergravitation bezeichnet wird, liefert klare
Vorhersagen f\"ur die Zerf\"alle des leichtesten
Standardmodell-Superpartners in das Gravitino und
Standardmodell-Teilchen. Diese Vorhersagen h\"angen nur von den
Massen der beteiligten Teilchen ab. Nach einer Messung des
Massen-Spektrums der Standardmodell-Superpartner an zuk\"unftigen
Beschleunigern wird die einzige Unbekannte in diesen Vorhersagen die
Gravitino-Masse sein.
Diese kann dann gefunden werden als der Wert, bei dem die gemessene
Lebensdauer des leichtesten Standardmodell-Superpartners mit der
entsprechenden theoretischen Vorhersage der Supergravitation
\"ubereinstimmt.
Da in den Zerf\"allen die Energie-Impuls-Erhaltung erf\"ullt sein
muss, ist zus\"atzlich auch eine Bestimmung der Gravitino-Masse aus
der Messung der Energien der anderen emittierten sichtbaren Teilchen
denkbar.
Mit der so bestimmten Gravitino-Masse sind die theoretischen
Vorhersagen eindeutig festgelegt, so dass eine m\"ogliche
\"Ubereinstimmung mit den Daten als experimenteller Nachweis f\"ur das
Gravitino als leichtester Superpartner und zugleich als Best\"atigung
der Supergravitation angesehen werden kann.
Eine Messung der Gravitino-Masse bestimmt dar\"uber hinaus die
Brechungsskala der Supersymmetrie und ist essenziell, um zu
entscheiden, ob das Gravitino tats\"achlich der fundamentale Baustein
der Dunklen Materie sein kann.

W\"ahrend ein langlebiger geladener leichtester
Standardmodell-Superpartner direkt nach seiner Produktion gut
unterscheidbar von den anderen an zuk\"unftigen Beschleunigern zu
erwartenden Reaktionen sein sollte, ist die Beobachtung eines solchen
Superpartners, der nach einer Produktion mit einer bereits geringen
Anfangsgeschwindigkeit im Detektor zur Ruhe kommt und dann dort zu
einem sp\"ateren Zeitpunkt zerf\"allt, als eine enorme Herausforderung
anzusehen.
Die nahezu fertiggestellten Detektoren am LHC sind z.B.\ entwickelt
worden, um sehr schnell Reaktionen aufzuzeichnen, und sind daher
weniger geeignet, Zerf\"alle langlebiger Teilchen zu untersuchen.
Insbesondere sind diese Detektoren auf Reaktionen ausgerichtet, die in
der N\"ahe des Punktes stattfinden, an dem die beschleunigten Teilchen
kollidieren.  Auf langsame langlebige geladene Superpartner, die im
Detektorvolumen mit einem relativ gro{\ss}en Abstand vom
Kollisionspunkt zu Ruhe kommen und dann sp\"ater zerfallen, sind diese
Detektoren und die dazugeh\"origen Analyseprogramme bisher nicht
vorbereitet.
Es gibt aber Arbeitsgruppen, in denen gegenw\"artig untersucht wird,
wie mit den Detektoren am LHC und an anderen zuk\"unftigen
Beschleunigern -- wie z.B.\ dem geplanten International Linear
Collider (ILC) -- Szenarien mit einem langlebigen geladenen
Superpartner optimal erforscht werden k\"onnen.
Hierbei wird auch \"uber zus\"atzliches Detektormaterial nachgedacht,
mit dem die Anzahl der gestoppten langlebigen Superpartner erh\"oht
werden kann (Abb.~\ref{Fig:4}d).
Gerade f\"ur die eindeutige Identifizierung des Gravitinos wird eine
hohe Anzahl von Zerf\"allen aufgezeichnet und analysiert werden
m\"ussen. 
F\"ur diese Identifizierung ist auch die Berechnung und Vorhersage
m\"oglicher Signaturen anderer extrem schwach wechselwirkender
Kandidaten f\"ur das leichteste Superteilchen, wie z.B.\ das so
genannte Axino, unerl\"asslich.  Da langlebige geladene
Standardmodell-Superpartner auch bei einer leichten Verletzung der
R-Parit\"at auftreten k\"onnen, m\"ussen auch solche Szenarien
entsprechend studiert werden.
Diese theoretischen Untersuchungen geh\"oren zu den
Forschungsaktivit\"aten am MPI f\"ur Physik.

\section{Zusammenfassung}

Kosmologische und astrophysikalische Untersuchungen zeigen, dass unser
Universum nur zu ca.\ 5\% aus den bisher entdeckten Teilchen besteht.
Ein wesentlich gr\"o{\ss}erer Teil von ca.\ 22\% des
Gesamtenergieinhaltes liegt in Form von Dunkler Materie vor. Da deren
fundamentaler Baustein nicht Teil des Standardmodells der
Elementarteilchenphysik sein kann, muss die Existenz der Dunklen
Materie als ein Hinweis auf neue Physik jenseits des Standardmodells
der Elementarteilchenphysik verstanden werden.
Die supersymmetrische Erweiterung des Standardmodells ist ein
besonders attraktives Konzept, da sie unter anderem eine elegante
L\"osung des Hierarchieproblems liefert und die Hypothese der Existenz
einer \"ubergeordneten, die Standardmodell-Kr\"afte
vereinheitlichenden Kraft untermauert.

Sollte die Supersymmetrie tats\"achlich in der Natur realisiert sein,
dann ist aufgrund der Proton-Stabilit\"at und der dadurch motivierten
Erhaltung der R-Parit\"at davon auszugehen, dass der leichteste
Superpartner stabil ist.
Damit ist als m\"oglicher leichtester Superpartner das leichteste
Neutralino, das nur der schwachen Kraft und der Gravitation
unterliegt, ein vielversprechender Dunkle-Materie-Kandidat.
Sollte das leichteste Neutralino tats\"achlich der fundamentale
Baustein der Dunklen Materie sein, dann sollte die schwache
Wechselwirkung des leichtesten Neutralinos ausreichen, diese in der
indirekten Suche, in der direkten Suche und an zuk\"unftigen
Beschleunigern nachweisen zu k\"onnen. Dieser Nachweis k\"onnte
bereits in den kommenden Jahren gelingen.

Ein weiterer sehr gut motivierter Kandidat f\"ur das leichteste
Superteilchen und die Dunkle Materie ist das Gravitino, das als der
Superpartner des Gravitons viel schw\"acher wechselwirkt als das
leichteste Neutralino. Sollte das Gravitino der fundamentale Baustein
der Dunklen Materie sein, dann wird man die Dunkle Materie aufgrund
der extrem schwachen Gravitino-Wechelwirkungen weder in der direkten
noch in der indirekten Suche nachweisen k\"onnen. Ein experimenteller
Nachweis kann dennoch an zuk\"unftigen Teilchenbeschleunigern
gelingen, wenn der n\"achstleichteste Superpartner ein elektrisch
geladenes Teilchen ist, dessen Zerf\"alle in das Gravitino noch im
Detektorvolumen beobachtet werden k\"onnen. Bereits in den n\"achsten
f\"unf Jahren k\"onnte ein langlebiger geladener leichtester
Standardmodell-Superpartner am LHC als erster Hinweis auf das
Gravitino als Baustein der Dunklen Materie gefunden werden.

Neben dem leichtesten Neutralino und dem Gravitino gibt es noch andere
Kandidaten f\"ur das leichteste supersymmetrische Teilchen wie z.B.\ 
das Axino, das wie das Gravitino trotz extrem schwacher
Wechselwirkungen ein vielversprechender Kandidat f\"ur die Dunkle
Materie ist. Sollten supersymmetrische Teilchen an zuk\"unftigen
Beschleunigern produziert werden k\"onnen, dann wird die
Identifikation des leichtesten Superpartners und damit
m\"oglicherweise die des fundamentalen Bausteins der Dunklen Materie
eine der zentralen Aufgaben sein.

Nach bisherigen Untersuchungen ist es vorstellbar, dass tats\"achlich
der Gro{\ss}teil der Materie in unserem Universum aus dem leichtesten
supersymmetrischen Teilchen besteht. Mit den kommenden Experimenten am
LHC und anderen zuk\"unftigen Beschleunigern kann es gelingen, genau
diesen fundamentalen Baustein der Dunklen Materie erstmals im Labor zu
produzieren und zu untersuchen. Die Entschl\"usselung der
teilchenphysikalischen Identit\"at der Dunklen Materie w\"urde eines
der gr\"o{\ss}ten R\"atsel der Naturwissenschaften l\"osen und zu den
gr\"o{\ss}ten Entdeckungen der Menschheit geh\"oren.

\end{document}